\documentclass{aa}
\usepackage{graphicx}

\def\nodata{ ~$\cdots$~ }

\begin{document}
%

\title{Stellar populations in NGC 5128 with the VLT: evidence for recent
star formation\thanks{Based on observations collected at the European
Southern Observatory, Paranal, Chile, within the Observing Programmes
63.N-0229 and 65.N-0164, and in part on observations collected by the
NASA/ESA Hubble Space Telescope, which is operated by AURA, Inc., under NASA
contract NAS5--26555.}}

\author{M. Rejkuba\inst{1,2}
    \and D. Minniti\inst{2}
    \and D.R. Silva\inst{1}
    \and T.R. Bedding\inst{3}}

\offprints{M. Rejkuba, E-mail: mrejkuba@eso.org}

\institute{European Southern Observatory, Karl-Schwartzschild-Strasse
           2, D-85748 Garching, Germany\\
           E-mail: mrejkuba@eso.org, dsilva@eso.org
  \and Department of Astronomy, P. Universidad Cat\'olica, Casilla
        306, Santiago 22, Chile\\
        E-mail: dante@astro.puc.cl
  \and School of Physics, University of Sydney 2006, Australia\\
        E-mail: bedding@physics.usyd.edu.au}

\date{Received date / Accepted date}
\authorrunning{Rejkuba et al.}
\titlerunning{Stellar populations in NGC 5128 with the VLT}

\abstract{We resolve stars of the nearest giant elliptical galaxy NGC~5128
using VLT with FORS1 and ISAAC.  We construct deep $U$, $V$ and $K_s$
color-magnitude and color-color diagrams in two different halo fields (in
the halo and in the north-eastern diffuse shell). In the outer, shell
field, at $\sim14$ kpc from the center of the galaxy, there is a
significant recent star formation with stars as young as 10 Myr,
approximately aligned with the prominent radio and x-ray jet from the
nucleus of the host AGN. Ionized gas filaments are evident in ultraviolet
images near the area where neutral H{\sc i} and CO molecular gas was
previously observed.  
The underlying stellar population of the halo of the
giant elliptical is predominantly old with a very broad metallicity
distribution. 
The presence of an extended giant branch reaching $M_{\rm bol}=-5$ mag
suggests the existence of a significant intermediate-age
AGB population in the halo of this galaxy.
}

\maketitle

%
%

\section{Introduction}

Galaxies show a wide range of star formation activity and a large range of
metallicities and ages in their stellar populations. 
Understanding the physical nature and origins of their
stars is fundamental to understanding the 
formation and evolution of galaxies.
The current knowledge of star formation histories of galaxies
along the Hubble sequence is summarized by Kennicutt~(\cite{kennicutt}). 
This knowledge is mainly based on the integrated
photometric and spectroscopic studies through the predictions of
populations synthesis of the integrated light of star clusters
(Bica~\cite{bica}) and galaxies (Bruzual \& Charlot~\cite{bc93}, 
Maraston~\cite{maraston}). The predictions of population synthesis can now 
be tested directly not only for globular clusters 
(e.g.\ Vazdekis et al.~\cite{vazdekis}), but also for 
nearby galaxies.

Stellar evolution theory provides predictions of the features
expected in color-magnitude diagrams (CMDs) for stellar populations with 
different ages and metallicities  (Renzini \& Fusi Pecci~\cite{renzini88},
Chiosi et al.~\cite{chiosi}, Gallart~\cite{gallart98}, 
Aparicio~\cite{aparicio98}, Tolstoy~\cite{tolstoy}). 
Coupled with the improvements in telescope sizes, detector sensitivity, 
field of view and spatial resolution, compared with those a decade ago, the 
direct observations of the stellar content of nearby galaxies 
is becoming one of the most active areas of extragalactic research.

In the Local Group, indicators of the old stellar populations such as old
main sequence turn-offs or at least the horizontal branch magnitudes are
within reach of available instrumentation (e.g.\  Hurley-Keller 
et al.~\cite{hurley-keller}, Gallart et al.~\cite{gallart99}, 
Held et al.~\cite{held},
Rejkuba et al.~\cite{rejkubaWLM}).
The recent results on the
studies of the Local Group galaxies have been summarized by
Mateo~(\cite{mateo}), van den Bergh~(\cite{vdb99,vdb00}) and
Grebel~(\cite{grebel}).

The Local Group contains galaxies representative of
almost all the classes, except the important giant elliptical class of
galaxies. The closest giant elliptical is 
NGC~5128, the dominant galaxy in the Centaurus group at distance of
$3.6 \pm 0.3$ Mpc (Harris et al.~\cite{harris99}, 
Soria et al.~\cite{soria}, Hui et al.~\cite{hui}, Tonry \&
Schechter~\cite{tonry}). 
It has been extensively studied over the last 50
years (for an exhaustive review see Israel~\cite{israel}). 
                                                                              
Its popularity is not only due to its brightness, but also to its many
unusual features: (i)~there is a prominent dust band containing young stars
and H{\sc ii} regions (Unger et al.~\cite{unger},
Wild \& Eckart~\cite{wild}, Graham~\cite{graham79}); 
(ii)~there is an active nucleus
with a radio and x-ray jet, radio lobes and optical filaments 
(Cooper et al.~\cite{cooper}, Feigelson et al.~\cite{feigelson},
Kraft et al.~\cite{kraft}, 
Schreier et al.~\cite{schreier}, Clarke et al.~\cite{clarke},
Blanco et al.~\cite{blanco}, Dufour \& van den Bergh~\cite{dufour}); 
(iii)~Malin et al.~(\cite{malin}) 
discovered a large number of faint narrow shells of stars 
surrounding the galaxy; (iv)~Schiminovich et al.~(\cite{schiminovich})
detected $4\times10^8$ M$_\odot$ of H{\sc i} gas associated with the stellar
shells, but slightly displaced outside the shells; and (v)~most recently, 
molecular CO gas has been found associated with the H{\sc i} gas and the stellar
shells (Charmandaris et al.~\cite{charmandaris}). All of these are
clear indications of a recent interaction with a gas--rich galaxy. 

The high resolution and sensitivity of WFPC2 on the Hubble Space Telescope
(HST) enabled the first studies of the resolved old stellar populations 
in the halo of NGC~5128 (Soria et al.~\cite{soria}, Harris 
et al.~\cite{harris99}, Harris \& Harris~\cite{harris00}, Mould 
et al.~\cite{mould00}). NICMOS on HST was used to resolve the stars in the
near IR (Marleau et al.~\cite{marleau00}) in the same field as the
optical study of Soria et al.~(\cite{soria}; $\sim$9kpc south of the 
center of the galaxy). In these two studies, a
small intermediate-age population of $\sim 5$Gyr 
has been found, comprising up to $\sim 10 \%$ of the total stellar population
in the halo. On the other hand, there are 
almost no intermediate--age stars in the field further out in the halo, at
$\sim 20$ kpc from the galaxy center (Harris et al.~\cite{harris99})
nor at $\sim 31$ kpc (Harris \& Harris~\cite{harris00}). 
The comparison of the two results may
indicate the presence of a gradient in the stellar population in the halo,
as suggested by Marleau et al.~(\cite{marleau00}). 
However, the small field of view of HST puts serious limitations to the 
conclusions 
in the cases where the strong gradients in galaxy populations exist (see, for
example, the case of Local Group dwarf galaxies like Leo~I (Gallart  
et al.~\cite{gallart99}, Held et al.~\cite{held}) or WLM (Minniti \&
Zijlstra~\cite{mz96,mz97})). 

We present here a wide wavelength range photometry of the resolved stellar
populations in NGC~5128 obtained from the ground with Very Large Telescope
(VLT) in $U$, $V$ and $K_s$ band. The deep and high-resolution VLT 
imaging, coupled with
the much larger field of view than HST, enables us to address
the questions of the gradients in stellar populations in the halo of this
giant elliptical galaxy. 
The proximity of NGC~5128 provides an unusual opportunity for a 
direct study of shell stars. We use infrared-optical colour-magnitude 
diagrams of the shell to study the ages and metallicities of the 
stars belonging the cannibalized galaxy.

%
%

 \section {The Data}
\label{data}
 
\begin{table*}
  \caption[]{Journal of Observations}
    \label{obslog}
      \begin{tabular}{llllllllll}
        \hline \hline
Field & $\alpha_{(2000)}$ & $\delta_{(2000)}$ & Date  
& Telescope \&    & Exposure & FILTER & Airmass&Seeing & Epoch   \\
\#    & (h min sec)       & ($^\circ$  $\arcmin$  $\arcsec$)&  dd/mm/yyyy 
& Instrument &  (sec)   &        &        & $\arcsec$ &  \\
\hline \hline

 1&13 26 23.5&$-$42 52 00&12/07/1999&Antu+FORS1  &2$\times$900&$U$ & 1.54 &0.52& \\ 
 1&13 26 23.5&$-$42 52 00&12/07/1999&Antu+FORS1  &2$\times$900&$V$ & 1.77 &0.54&\\ 
 1&13 26 23.5&$-$42 52 00&08/04/1999&Antu+ISAAC  &10$\times$36$\times$10& $K_s$& 1.15 &0.41&A\\
 1&13 26 23.5&$-$42 52 00&07/05/1999&Antu+ISAAC  &10$\times$36$\times$10& $K_s$& 1.05 &0.40&B \\
 1&13 26 23.5&$-$42 52 00&28/05/1999&Antu+ISAAC  &10$\times$36$\times$10& $K_s$& 1.14 &0.51&C\\
 1&13 26 23.5&$-$42 52 00&15/04/2000&Antu+ISAAC  &65$\times$6$\times$10& $K_s$& 1.09 &0.35&D\\
 1&13 26 23.5&$-$42 52 00&10/05/2000&Antu+ISAAC  &65$\times$6$\times$10& $K_s$& 1.22 &0.43&E\\
 1&13 26 23.5&$-$42 52 00&07/06/2000&Antu+ISAAC  &65$\times$6$\times$10& $K_s$& 1.13 &0.60&F\\
 1&13 26 23.5&$-$42 52 00&08/07/2000&Antu+ISAAC  &65$\times$6$\times$10& $K_s$& 1.15 &0.31&G\\
 2&13 25 24.0&$-$43 10 00&11/07/1999&Antu+FORS1  &2$\times$900&$U$ & 1.28 & 0.53&\\ 
 2&13 25 24.0&$-$43 10 00&11/07/1999&Antu+FORS1  &2$\times$900&$V$ & 1.20 & 0.44&\\
 2&13 25 24.0&$-$43 10 00&08/04/1999&Antu+ISAAC  &10$\times$36$\times$10& $K_s$& 1.05 &0.36&A\\
 2&13 25 24.0&$-$43 10 00&07/05/1999&Antu+ISAAC  &10$\times$36$\times$10& $K_s$& 1.10 &0.39&B\\
 2&13 25 24.0&$-$43 10 00&29/05/1999&Antu+ISAAC  &10$\times$36$\times$10& $K_s$& 1.06 &0.44&C\\
 2&13 25 24.0&$-$43 10 00&15/04/2000&Antu+ISAAC  &65$\times$6$\times$10& $K_s$& 1.09 &0.40&D\\
 2&13 25 24.0&$-$43 10 00&12/05/2000&Antu+ISAAC  &51$\times$6$\times$10& $K_s$& 1.22 &0.51&Eb\\
 2&13 25 24.0&$-$43 10 00&19/05/2000&Antu+ISAAC  &20$\times$6$\times$10& $K_s$& 1.22 &0.44&Ec\\
 2&13 25 24.0&$-$43 10 00&07/06/2000&Antu+ISAAC  &41$\times$6$\times$10& $K_s$& 1.13 &0.57&Fa\\
 2&13 25 24.0&$-$43 10 00&09/06/2000&Antu+ISAAC  &30$\times$6$\times$10& $K_s$& 1.13 &0.40&Fb\\
 2&13 25 24.0&$-$43 10 00&08/07/2000&Antu+ISAAC  &65$\times$6$\times$10& $K_s$& 1.15 &0.40&G\\
 2&13 25 24.0&$-$43 09 04&20/02/2000&NTT+SOFI    &45$\times$10$\times$6& $K_s$& 1.04 &0.55&\\
 2&13 25 24.3&$-$43 09 58&27/06/1998&HST+NIC3    &256                  & F222M&      &    &\\
\hline
        \end{tabular}
\end{table*}

\begin{figure*}
	\vspace{17cm}
   \caption[]{Combined $U$- and $V$- band image of Field~1, taken at the VLT
(UT1+FORS1). The field of view is $6\farcm8 \times 6\farcm8$. North is at
the top and east to the left.}
   \label{f1}
\end{figure*}

\begin{figure*}
	\vspace{17cm}
   \caption[]{Combined $U$- and $V$- band image of Field~2, taken at the VLT  
(UT1+FORS1). The field of view is $6\farcm8 \times 6\farcm8$. North is at
the top and east to the left.}
   \label{f2}   
\end{figure*}

The observations were carried out with the ESO Very Large 
Telescope (VLT)  at Paranal Observatory, Chile. They 
consist of optical (Bessel $U$- and $V$-band) and near-IR 
($K_s$-band) images. We observed two fields in the halo of 
NGC~5128. Field~1 ($\alpha_{2000}=13^h26^m23.5^s$,
$\delta_{2000}=-42^{\circ}52\arcmin00\arcsec$; Fig.~\ref{f1})
was centered on the prominent N-E shell, $\sim14\arcmin$ away from the
center of the galaxy. Field~2 ($\alpha_{2000}=13^h25^m26^s$,
$\delta_{2000}=-43^{\circ}10\arcmin00\arcsec$; Fig.~\ref{f2})
was chosen to overlap with the HST field of
Soria et al.~(\cite{soria}) and lies at a
distance of $\sim9\arcmin$ from the center of the galaxy.
The optical data were obtained on 1999 July 11 and 12, 
while the $K_s$-band 
images cover several epochs and will be used
to search for long-period variable stars.
All of the observations were secured in the service mode. Calibration data, 
bias, dark and flat-field images as well as  
photometric standards from Landolt (in optical; \cite{landolt}) 
and Persson et al. (IR; \cite{persson}) catalogues,
were supplied by the ESO calibration plan. The observations are summarized in 
Table~\ref{obslog}. 

\subsection{Optical Observations and Data Reduction}

Two 15-minute exposures in both $U$ and $V$ were acquired for each of the 
two fields in NGC~5128 
using FORS1 (FOcal Reducer and low dispersion Spectrograph)
on VLT Unit Telescope~1 (Antu).
The FORS1 detector is a
2048$\times$2048 Tektronix CCD, thinned and anti-reflection coated. The
pixel size is 24$\times$24 $\mu$m.
The field of view is $6\farcm8 \times 6\farcm8$ and
the scale is $0\farcs2/$pixel. 

For service observations in direct-imaging mode, the FORS1 CCD is read out 
in four-port read-out mode. We used the ESO pipeline reductions 
which are based on MIDAS package specially developed to
reduce images with 4 different amplifiers. With it the overscan for
each amplifier was subtracted individually and after the subtraction of bias
the images were corrected for the flat-field.

\subsection{Near-IR Observations and Data Reduction}

For the near-IR $K_s$-band observations we used ISAAC, also on Antu. In this 
wavelength domain (0.9--2.5 $\mu$m) the detector is a 
1024$\times$1024 Hawaii Rockwell array. 
The field of view is $2\farcm5 \times 2\farcm5$ and
the scale is $0\farcs147/$pixel.  A series of 10-second exposures were
taken at each epoch, usually in groups of six, with the number of repeats 
depending on weather
conditions and any technical problems. The total exposure time for each
epoch is given in Table~\ref{obslog}.

An additional $K_s$-band observation of Field~2, with a total on-target
exposure time of 45 minutes, was acquired with the SOFI instrument at 
the ESO 3.5-m New Technology Telescope (NTT) at La Silla Observatory, Chile. 
The field of view of SOFI is $4\farcm94 \times 4\farcm94$ 
and the scale is $0\farcs292/$pixel. 

We also obtained observations with the NIC3 array on HST, using the F222M
filter, which is similar to (although narrower than) the standard K filter.
The NIC3 field of view is $51\farcs2 \times 51\farcs2$, much smaller than
the one of ISAAC.

The standard procedure in reducing IR data consists of (i) dark 
subtraction, (ii) flat-field correction, (iii) sky subtraction, (iv) 
registering and combining the images.
We did not use the ESO pipeline reduction product.
Good sky subtraction in a crowded field like that of a galactic halo is
particularly important. For that step we used the
DIMSUM package (Stanford et al.~\cite{dimsum}) within IRAF\footnote{IRAF is
distributed by the National Optical Astronomy Observatories, which is
operated by the Association of Universities for Research in Astronomy, Inc.,
under contract with the National Science Foundation}.
In DIMSUM the sky subtraction is made in two passes. In the first, a median
sky is computed for each image from the six frames that are closest in time.
The shifts between the sky-subtracted frames are then computed and all the
images stacked together using a rejection algorithm to remove cosmic rays.
An object mask is computed for the coadded image and then shifted back in
order to create object masks for the individual frames. In the second pass,
the sky subtraction is made using the object masks to avoid overestimation of
the sky level. These masks are also used to check that the bright object
cores were not removed as cosmic rays in the previous pass.
After the mask-pass sky subtraction, all frames belonging to single epoch are
registered with {\em imalign} and combined with {\em imcombine} task in IRAF.

\section{The Photometry}

\subsection{Data analysis}

The photometric reduction of the combined $U$-, $V$- and 
$K_s$-band images was performed
using the DAOPHOT~II programme (Stetson~\cite{stetson, stetsonALF}). 
First, we located all the objects that were $>3 \sigma$ above the 
background on individual images. 
More than 50 relatively bright, not saturated, isolated, 
stellar objects were chosen to create the variable PSF for each image. 
ALLSTAR fitting of the PSF to all the objects produced the object
lists that were matched with DAOMATCH and DAOMASTER, where only objects
with good photometry in at least two frames were kept. The final
photometric catalogue was obtained with ALLFRAME which uses as
input information photometry lists from ALLSTAR and fits the PSF to all
frames (in $U$, $V$ and $K$-band) simultaneously. 
Again, only the objects detected in at least 2 images
were kept. Using the information on the location of stars in all bands
simultaneously improved our photometry, which is deeper by $\sim1$ mag with
respect to ALLSTAR photometry. Moreover, the treatment of close companions,
in particular the ones 
that have different colors, is much better with ALLFRAME.

NGC~5128 is close enough that its globular clusters appear slightly
resolved, in the sense of having a larger FWHM and non-stellar PSF
(Minniti et al.~\cite{mi+96}, Rejkuba~\cite{rejkuba01}). 
Restricting the sharpness and goodness of the fit parameters to 
$-0.7<$SHARP$<0.7$, we rejected most of the
galaxies, star clusters and other extended objects as well as remaining
blemishes and cosmic rays from the final photometry list. Also stars with
large photometric uncertainties in one or more filters ($\sigma \ge 0.5$ mag)
were rejected.
Finally the images were visually checked and 
a few remaining extended background 
objects (e.g. partially resolved star forming region in a background 
spiral galaxy)  were discarded.
With this selection our final $U,V$ photometry catalogue contains 1581 and 
1944 stars in Fields~1 and 2, respectively, while the $V,K$ catalogue contains
5172 and 8005 stars and the numbers of stars with good photometry in $U$, $V$
and $K$-band are 508 and 663 in Fields~1 and 2, respectively.

\subsection{The photometric calibration}

\begin{figure}
   \resizebox{\hsize}{!}{\includegraphics[angle=270]{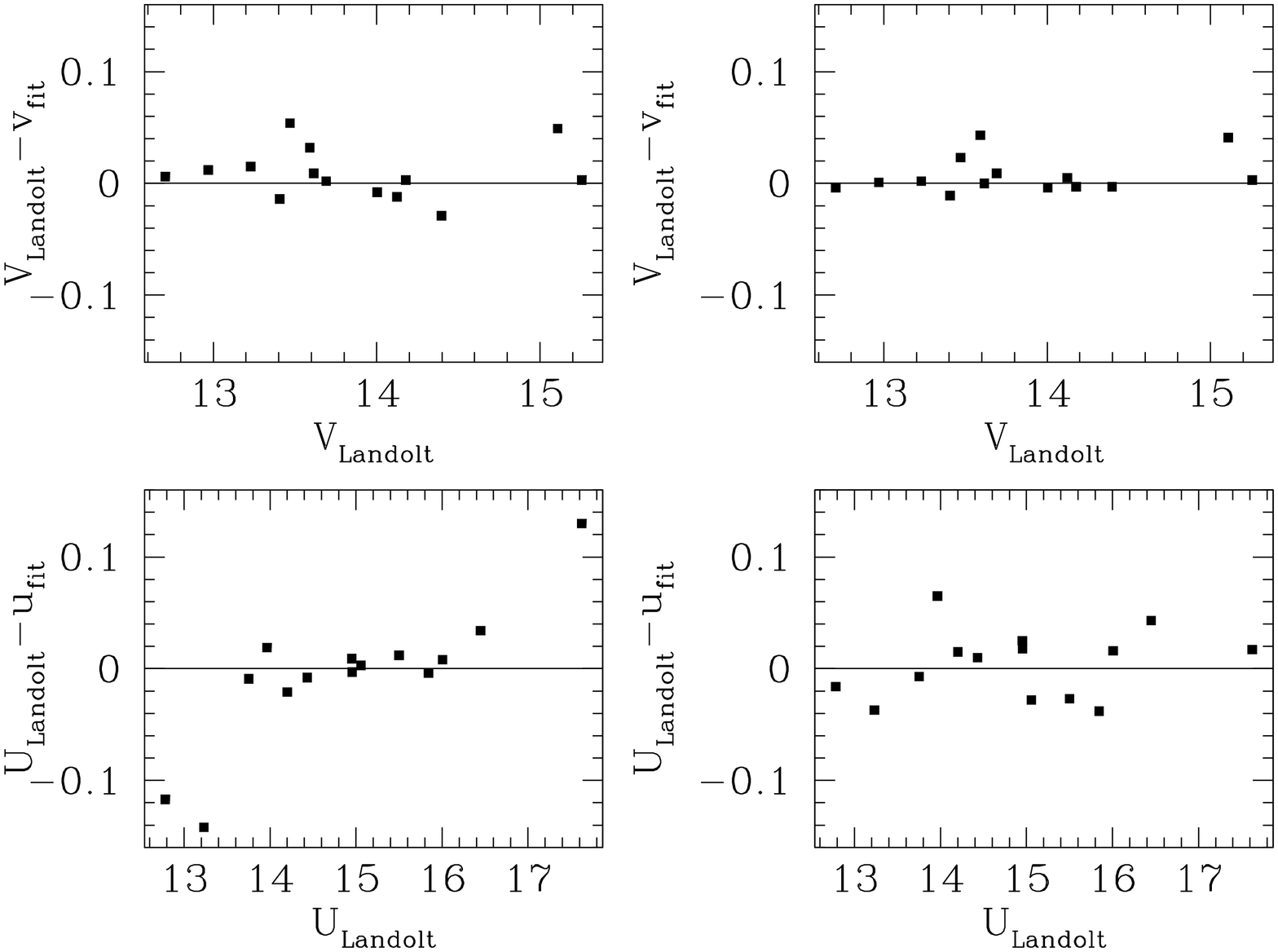}}
   \caption[]{Photometric calibration of Landolt (\cite{landolt}) stars during
the nights of observations in $U$ and $V$ filters. Left panels:
the scatter for the calibration without the color term;
right panels: the scatter for the calibration with the color term.}
   \label{calib}
\end{figure}

For the photometric calibration of the optical images, standard stars from 
the catalogue of Landolt~(\cite{landolt}) were used. 
We checked the photometric 
quality of the two nights separately and since  both were photometric,
with almost identical zeropoints, we
combined the standard stars for the two nights. Using a total of 14 stars in
4 different fields, spanning the color range $-1.321<(U-V)< 4.162$, 
we derived the following calibration transformations:
\begin{eqnarray}  
u_{\rm inst} & = & U - 24.262(\pm0.089 ) + 0.379 (\pm0.068 )*X \nonumber \\
         &   & -0.042(\pm0.007)*(U-V)
\label{Utran}
\end{eqnarray}

\begin{equation}
v_{\rm inst} = V - 27.348 (\pm 0.042)  + 0.213 (\pm 0.034)*X
\label{Vtran}
\end{equation}
where $X$ is the mean airmass of the observations, $u_{\rm inst}$ and $v_{\rm
inst}$ are
instrumental magnitudes and $U$ and $V$ magnitudes from the Landolt
(\cite{landolt}) catalogue.
The one-sigma scatter around the mean was 0.031 mag for the
$U$ band and 0.023 mag
for the $V$ band (Fig.~\ref{calib}, left panel).
Adding the color term ($U-V$) in the transformations slightly
reduces the scatter in the $V$-band to 0.017 mag (Fig.~\ref{calib},
right panel). However, the calibration equation without the color term for
the $V$-band was preferred, since the scientific data in that 
filter go much deeper and $U$
magnitudes for some objects could not be measured accurately enough.

Observations at our reference $K_s$-band epoch (G in Table~\ref{obslog}) were taken
in photometric conditions. During the same night, July 8 2000,  
three standard stars from the list of Persson et
al.~(\cite{persson}) were observed. Each standard star
was observed at 5 different positions on the IR-array. In this way, a total of
15 independent measurements were obtained. However, the number of
measurements with different airmass was only 3, so that we preferred to adopt
the mean extinction coefficients measured on Paranal for $K_s$~-~band 
of 0.05 mag/airmass. 
The derived zeropoint of the
G-epoch observations is $24.23\pm 0.04$. The following calibration equation
was applied to our data:
\begin{equation}
k_{s,\rm inst} = K_s - 24.23 (\pm 0.04) + 0.05 * X
\label{Ktran}
\end{equation}
where $X$ is the airmass of observations, $k_{s,\rm inst}$ is the instrumental
magnitude and $K_s$ is the calibrated magnitude. The photometry of all
other $K$-band 
epochs were measured with respect to the reference epoch.

\subsection{Completeness and contamination}
\label{compl_cont_section}

\begin{figure}
\centering
\includegraphics[width=6.8cm,angle=270]{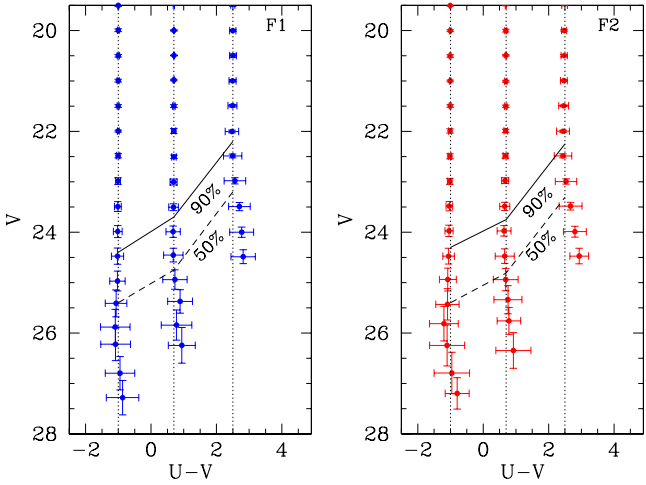}
  \caption[]{The completeness and photometric uncertainties as a 
function of magnitude and color. The thin vertical dotted lines indicate the
value of input $U-V$ colors. The full line defines the 90\% 
completeness limit and the dashed line is for 50\% completeness limit.} 
  \label{mag_complUV}
\end{figure}

\begin{figure}
\centering
\includegraphics[width=6.8cm,angle=270]{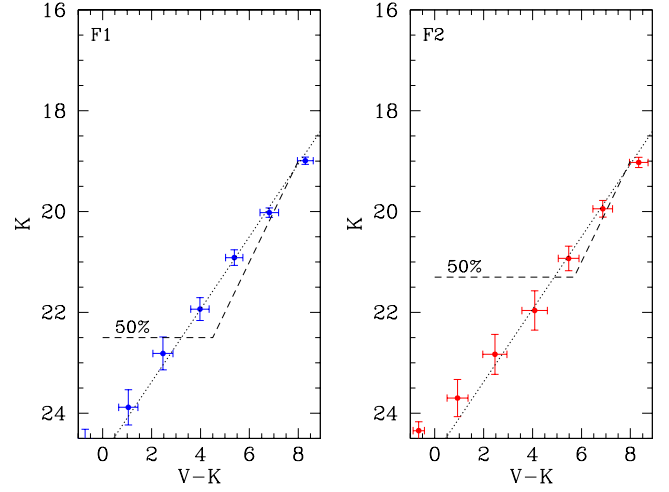}
  \caption[]{The completeness and photometric uncertainties as a 
function of magnitude and color. The thin dotted line indicate the
value of input $V-K$ colors. The dashed line defines the 50\% 
completeness limit.}
  \label{mag_complVK}
\end{figure}

\begin{figure}
\centering
\includegraphics[width=6.8cm,angle=270]{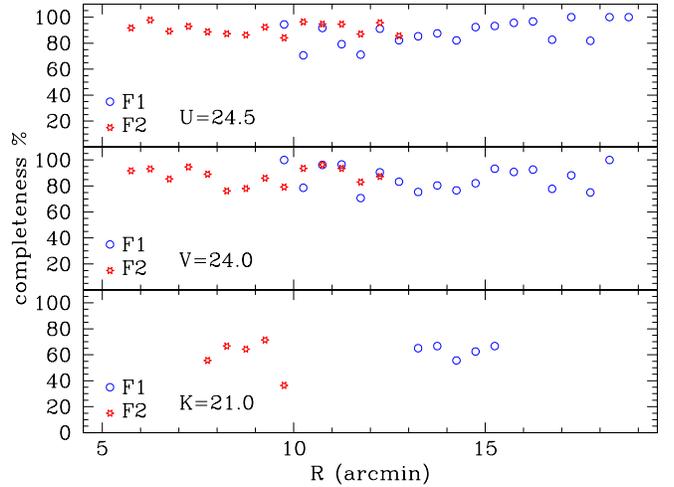}   
  \caption[]{The completeness as function of radial distance from the center
of NGC~5128 calculated around 90\% completeness limit for $U$- and $V$-band
(for $(U-V)=0.7$) 
and around 65\% for $K_s$-band.}     
  \label{rcomplUVK}
\end{figure}

We made extensive tests to measure completeness and magnitude uncertainties 
as a function of magnitude and radial distance from the center of the galaxy. 
The completeness for the $U$, $V$ and $K_s$-band photometry has 
been calculated using the ADDSTAR programme within DAOPHOT. 
We made twenty artificial star experiments, adding each time $\sim3000$
stars to the first frame. The stars were added on a regular grid separated
by $2\times R_{PSF} +1$ pixels in order not to
alter crowding (where $R_{PSF}$ is the PSF radius used for fitting the 
image with the worst seeing)  and having magnitudes randomly distributed in 
the observed range. The position of the first star in the list was chosen
randomly, so that over 20 different experiments the added stars were uniformly
distributed over the whole field. After the appropriate 
coordinate shifts were applied and magnitudes changed to instrumental
system taking into account the observed magnitudes and colors, the same
stars from the first frame were added also to all other frames. 
Their photometry was 
recomputed in the same way as for the original images. The stellar PSF
obtained from the field stars for the respective image was used in 
the simulations. Incompleteness in Field~1, 
defined by a recovery rate of 50\% from the artificial-star experiment,  
sets in around magnitude 
25 in the $U$-band (but depends strongly on $U-V$
color; see Fig.~\ref{mag_complUV}) 
and 22.5 in $K_s$-band (Fig.~\ref{mag_complVK}). 
The corresponding numbers for Field~2 are 25 for $U$-band
and 21.3 for $K_s$-band. 
The difference at
$K_s$-band between the two fields is due to (1)~better
seeing in Field~1 ($FWHM=2.1$ pix vs. 2.7 pix) and (2)~higher
surface brightness in Field~2.

Dependence of the completeness on radial distance from the center of the
galaxy ($\alpha_{2000}=13^{\rm h}25^{\rm m}26\fs4$, 
$\delta_{2000}=-43^\circ 01\arcmin 05\farcs1$)
was calculated for the magnitude bin around 90\%
level of completeness in $U$- and $V$-band (Fig.~\ref{rcomplUVK}) and around
65\% of completeness in $K_s$-band. 
There is no significant spatial variation of completeness in our data.

To assess the accuracy of our photometry, we calculated the difference between
the input and recovered magnitudes for each magnitude bin
(Fig.~\ref{mag_complUV}, \ref{mag_complVK}). 
Our photometry is reliable down to the incompleteness limit and 
blending is not seriously affecting our data (see the discussion in
Sec.~\ref{CMD_VK}).
For magnitudes fainter than the 50\%
completeness limit, the measured values are systematically brighter, 
because of the
bias towards brighter fluctuations and blending due to crowding.
The colors of the recovered stars with magnitudes fainter than the 50\% 
completeness limits are redder than the input colors 
in the $UV$ CMDs (Fig.~\ref{mag_complUV}), 
due to the larger incompleteness in the $U$- than in the $V$-band. In the
$VK$ CMDs
the colors of the recovered stars range from slightly 
redder than the input color for
the very red stars, due to the dominant incompleteness in the $V$-band, to
bluer for the faintest and bluest stars, due to the dominant incompleteness
in the $K$-band (Fig.~\ref{mag_complVK}). We did not 
correct our data for this systematic shift, because magnitudes fainter than
the 50\% completeness limit will not be used in further analysis.

\begin{figure}
\centering
\includegraphics[width=6.8cm,angle=0]{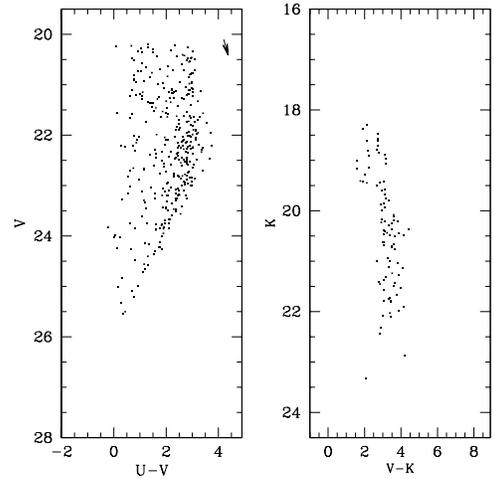}   
  \caption[]{$UV$ (left) and $VK$ (right) color-magnitude diagrams 
for foreground Galactic stars simulated using the Besan\c{c}on group model. 
The correction for completeness and realistic photometric uncertainties have been 
applied by adding the simulated stars to our Field~2 images and re-measuring 
them. The reddening vector $E(B-V)=0.1$ is plotted in the upper right corner.}
  \label{besancon}
\end{figure}

Contamination by foreground Galactic stars and unresolved background galaxies
is important because of the low Galactic latitude of our fields 
($b=19\fdg5$) and the very deep photometric limits observed. 
We used the Besan\c{c}on group model of stellar population synthesis of the 
Galaxy available through the 
Web\footnote{http://www.obs-besancon.fr/www/modele/modele\_ang.html}  
(Robin \& Creze~\cite{robin&creze}, 
Robin et al.~\cite{robin}) to simulate the total number 
and optical magnitude and color distribution of Galactic foreground 
stars in our fields. 
The simulated catalogue has 1827 stars in the FORS1 field 
($6\farcm8\times6\farcm8$) in the magnitude interval 
$18<V<30$. In order to get a realistic estimate for the number of stars
that would be observed in our fields, the correction for completeness 
is necessary. Therefore, we
added the stars from the simulated catalogue (with magnitudes scaled to
correspond to instrumental magnitudes) to our images and re-measured
their magnitudes. In this way, realistic photometric uncertainties were
applied and
the number of stars recovered in two fields was corrected for completeness.
A total of 340 and 350 stars with $U$- and $V$-band photometry satisfying
profile fitting and photometric uncertainty 
selection criteria were measured in Field~1 and 2, respectively.
Most of the foreground stars have $0<U-V<3$ (Fig.~\ref{besancon}). All
of the stars in the red part of the Field~2 CMD brighter than $V\sim22$ 
are expected to be foreground stars (see Sect.~\ref{CMD_UV}).
In order to adjust for the expected number of Galactic stars we 
normalized the models to the observed number of reddest stars in
Field~2. Thus the total number of foreground stars was increased by 31\% 
in both fields.

The Galaxy model simulation supplies  
not only the colors of the simulated stars, but also their metallicities, 
ages, spectral types and luminosity classes. Using all these data and 
the Kurucz (\cite{kurucz}) 
model atmospheres\footnote{http://cfaku5.harvard.edu/grids.html}, 
we derived the foreground contamination in the $K$-band.
After correction for completeness, the expected number of foreground stars 
in ISAAC  images is 112 and 91 in the $VK$ CMD of Field~1 and 2, respectively.
All of the foreground stars have $1.0< V-K < 4.5$
(Fig.~\ref{besancon} right panel). 

The measured number of compact 
background galaxies on the FORS1 images, taken in
similar observing conditions, is $\la 400$ in magnitude 
range $V=20-25$ mag for the selection of sharpness parameter $-1<sharp<1$ 
(Jerjen \& Rejkuba~\cite{jr01}). Our tighter selection
criteria ($-0.7<sharp<0.7$) eliminated most of them. 
In the smaller field of view of ISAAC, the predicted
number of background galaxies is $\sim330$ in the interval of
magnitudes $16<K_s<23$ and $\sim180$ between $16<K_s<22$
(Saracco et al.~\cite{saracco}). 
Most of the background galaxies are resolved and rejected by sharp and
magnitude uncertainty parameter requirements on our photometry. 
Only few compact galaxies might contaminate the sample. 

\subsection{Comparison with published data}

\begin{table}
  \caption[]{Comparison with Fasset \& Graham (FG) photometry; the numbers
of stars in the first column are from FG}
    \label{compare_phot_tab}
      \begin{tabular}{rccccc}
        \hline \hline
\#& $V_{\rm FG}$ & $(B-V)_{\rm FG}$ & $(U-B)_{\rm FG}$ & $V_{\rm FORS1}$ & $U_{\rm FORS1}$\\
\hline \hline

 1           & 19.94&   +0.23& $-$0.54& 19.987 & 19.498 \\
 4\rlap{$^a$}& 21.07&   +0.12& $-$1.26& 20.871 & 19.711 \\
 5\rlap{$^b$}& 21.19&   +0.04& $-$1.26& 20.925 & 19.717 \\
 7\rlap{$^c$}& 21.51& $-$0.07& $-$0.34& \nodata& 20.291 \\
10           & 21.65&   +0.04& $-$1.16& 20.986 & 20.440 \\
12           & 21.72&   +0.21& $-$0.20& 21.722 & 21.580 \\
13           & 21.75&   +0.11& $-$0.84& 21.667 & 20.600 \\
15           & 21.86&   +0.09& $-$0.61& 21.885 & 21.099 \\
16           & 21.94& $-$0.05& $-$0.92& 21.978 & 20.862 \\
17           & 21.95&   +0.34& $-$0.82& 21.963 & 21.391 \\
18\rlap{$^d$}& 22.01&   +0.30& $-$0.84& \nodata& \nodata\\
20           & 22.10&   +0.07& $-$0.82& 22.150 & 21.144 \\
21           & 22.12& $-$0.08& $-$1.00& 22.127 & 20.942 \\
26           & 22.25&   +0.08& $-$0.98& 22.323 & 21.234 \\
27           & 22.35& $-$0.03& $-$0.91& 22.331 & 21.295 \\
28           & 22.40& $-$0.05& $-$0.87& 22.412 & 21.323 \\
30           & 22.51&   +0.20& $-$0.56& 22.631 & 22.032 \\
32           & 22.53&   +0.13& $-$0.50& 22.579 & 22.217 \\
33           & 22.54&   +0.07& $-$0.76& 22.623 & 21.528 \\
35           & 22.55&   +0.11& $-$0.55& 22.503 & 21.968 \\
36           & 22.58& $-$0.08& $-$0.88& 22.606 & 21.535 \\
37           & 22.58&   +0.02& $-$1.06& 22.777 & 21.549 \\
39           & 22.69& $-$0.12& $-$0.82& 22.706 & 21.585 \\
42           & 22.82&   +0.00& $-$1.02& 22.856 & 21.646 \\
46           & 22.89&   +0.31& $-$0.12& 22.919 & 23.260 \\
47           & 22.90& $-$0.05& $-$1.00& 22.934 & 21.737 \\
48           & 22.96&   +0.08& $-$0.74& 23.051 & 22.149 \\
50           & 22.99& $-$0.04& $-$0.81& 23.134 & 22.027 \\
\hline
        \end{tabular}
\smallskip\\
{$^a$ f1.GC-8 (Rejkuba~\cite{rejkuba01})}\\
{$^b$ f1.GC-25 (Rejkuba~\cite{rejkuba01})}\\
{$^c$ saturated in FORS1 $V$ image}\\
{$^d$ extended in FORS1 images}
\end{table}

\begin{figure}
\centering
\includegraphics[width=8.8cm,angle=0]{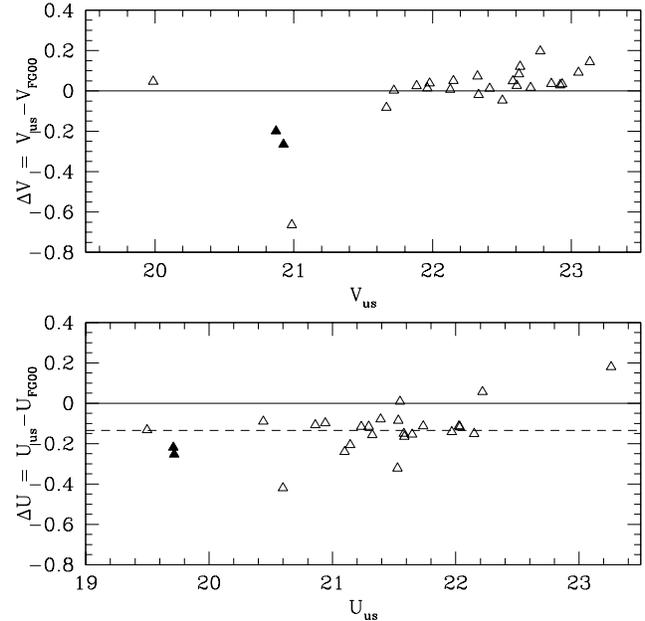}   
  \caption[]{Comparison of our photometry with that of Fasset \&
Graham~(\cite{fg00}) for 26 bright blue stars in common. 
In the upper panned
the comparison of $V$ band magnitudes is presented and in the lower panel
U-band magnitudes are compared. The two extended objects (\#4 and \#5 in
Tab.~\ref{compare_phot_tab}) noted by
Rejkuba~(\cite{rejkuba01}) are plotted with filled symbols.
The mean offset of -0.13 mag in $U$-band is
indicated with the dashed line.
}
  \label{compare_photo}
\end{figure}

Several recent studies of resolved stellar populations in NGC~5128 
exist in the literature, all but one made with HST in F606W ($V$) and 
F814W ($I$) or F110W ($J$) and F160W ($H$)  
photometric bands, which are not very sensitive to the recent star
formation. Our Field~2 is centered on the field of Soria et al.~(\cite{soria}),
which was also
observed in the near-IR (F110W and F160W filters) by Marleau et
al.~(\cite{marleau00}), but
the direct comparison is not possible due to different photometric bands
used. 

The only possible direct comparison is for our Field~1 photometry, which
partially overlaps with the HST
photometry of Mould et al.~(\cite{mould00}) and the ground based photometry of 
Fasset \& Graham~(\cite{fg00}). The last authors observed a wider field than
ours in $U$, $B$ and $V$ Glass filters, but due to smaller telescope aperture
(2.5m), their photometry is much shallower. The HST photometry is
deeper, but covers a much smaller area. The mean difference between 
$V$ magnitudes of Fasset \& Graham and Mould et al.\ photometry is $-0.13\pm
0.07$ mag, in the sense of HST photometry having a systematically fainter 
zero point (Fasset \& Graham \cite{fg00}). Comparison of 26 stars in common
between our data and that of Fasset \& Graham for the brightest blue stars
(their Table~3) is presented in Fig.~\ref{compare_photo}. The mean
difference for all 26 stars is negligible for $V$-band photometry, but it
amounts to 0.13 mag in $U$-band. 
{}From Fig.~\ref{compare_photo} a systematic trend with the magnitude is
apparent in $V$ and $U$ band. Excluding the star \#10 (see
Tab.~\ref{compare_phot_tab}), which has HST $V$ magnitude of 21.04 (thus 
close to our value; star R4 in Table~1 of Mould et al.), and extended
sources \#4, \#5 (see Rejkuba~\cite{rejkuba01}; filled triangles in
Fig.~\ref{compare_photo}) and \#18, 
these trends can be represented by the following equations:
\begin{equation}
V_{\rm FORS1}-V_{\rm FG} = 0.030 \times V_{\rm FORS1} - 0.65
\end{equation}
\begin{equation}
U_{\rm FORS1}-U_{\rm FG} = 0.089 \times U_{\rm FORS1} - 2.05   
\end{equation} 
with rms equal to 0.055 for $V$ and 0.096 for $U$.
At least part of this difference may be due to different filters 
used (Bessell~\cite{bessel}). Fasset \& Graham further neglected the color
term in their calibration due to insufficient color coverage of their
standards, while we found that the color term is important for our $U$-band
calibration. 

%
%

\section{The Color-Magnitude Diagrams}

\subsection{Optical CMDs}
\label{CMD_UV}

\begin{figure*}
\centering
\includegraphics[width=12.8cm,angle=0]{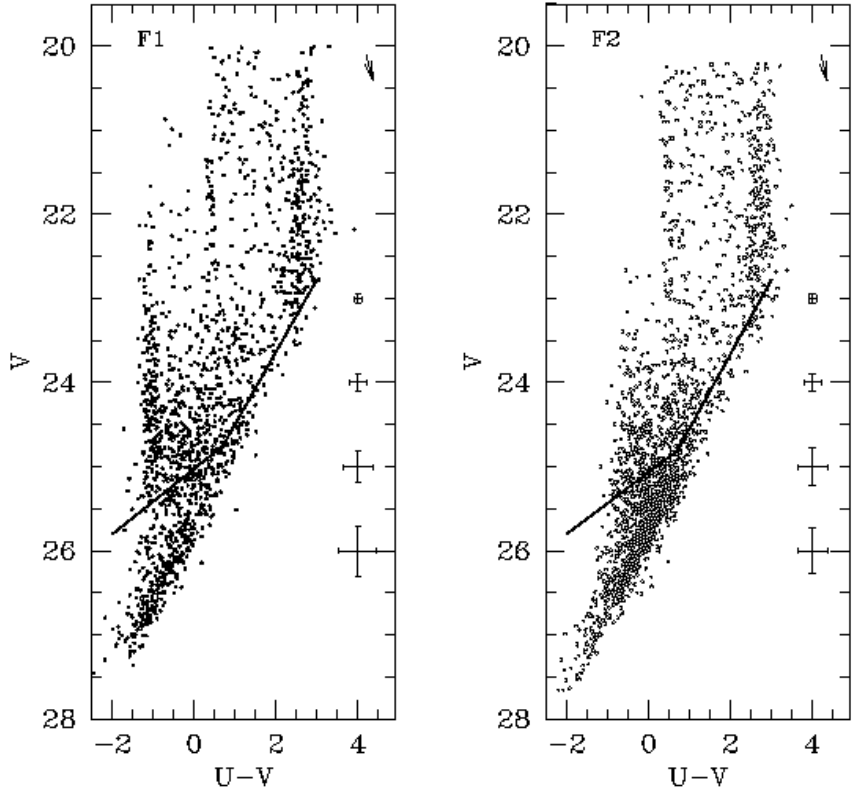}   
  \caption[]{UV color-magnitude diagram for Field~1 (left) and Field~2
(right). The arrow represents the reddening vector of $E(B-V)=0.1$. Typical
photometry uncertainties are plotted on the right side. The solid line
indicates the 50\% completeness.}
  \label{UVcmd}
\end{figure*}

Optical CMDs probe young and intermediate age populations. Theoretically,
the $U$ and $V$ light is dominated by young main sequence stars
(Buzzoni~\cite{buzzoni}).

Figure~\ref{UVcmd} shows ($U-V$) -- $V$ color-magnitude diagrams for stars
in both observed halo fields of NGC~5128. Due to better seeing  
the saturation magnitude of the $V$-band is $\sim0.3$ mag fainter 
for Field~2. Most of the stars redder than $(U-V)\sim0$ belong to our
own Galaxy (see Fig.~\ref{besancon}). 
The most important characteristic of the $UV$ CMD of Field~1 is
the upper main sequence, visible as the blue plume at $(U-V)\sim-1$ mag. 
By contrast, there are no such young massive stars in Field~2. 

\begin{figure*}
\centering
\includegraphics[width=12.8cm,angle=0]{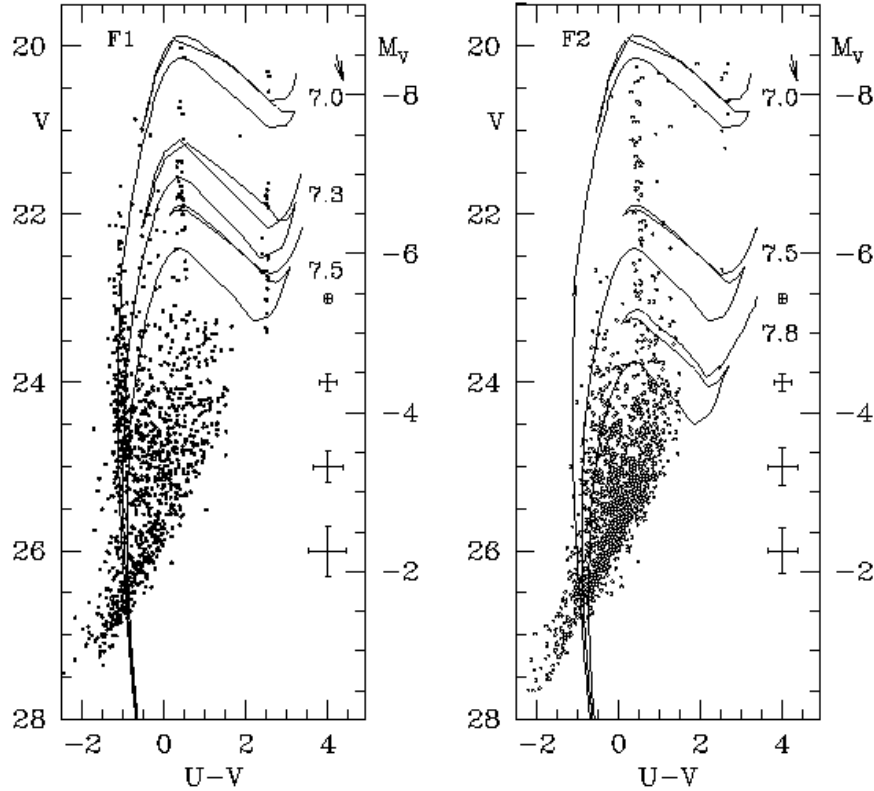}   
  \caption[]{CMDs with foreground stars statistically subtracted.
Overplotted are isochrones from Bertelli et al.~(\cite{bertelli94}) for
$Z=0.004$ with 
log(age) indicated with the number on the right side. The size of the
reddening vector corresponds to $E(B-V)=0.1$ mag.}
  \label{UVcmd_clean}
\end{figure*}

The Besan\c{c}on Galaxy model  (see previous section) was used to
``clean'' the CMDs of foreground star contamination. 
In Fig.~\ref{UVcmd_clean} we show $UV$ CMDs after the subtraction of 
foreground stars.
Overplotted are isochrones from Bertelli et al.~(\cite{bertelli94}) 
for $Z=0.004$ and for log(age)=7.0, 7.3, 7.5 for Field~1 and
log(age)=7.0, 7.5 and 7.8 for Field~2. The isochrones were shifted to the
distance of NGC~5128, assuming a distance modulus of $(m-M)_V=27.8$ and
reddening of $E(B-V)=0.11$ mag (Schlegel et al.~\cite{schlegel}). 
The difference between the two diagrams is
striking: the blue main sequence containing stars as young as $\sim10$ Myr
that is present in Field~1 is completely absent from the CMD of Field~2.
Well separated from the main sequence in Field~1 
is the sequence of blue core-helium
burning (BHeB) stars at $(U-V)\sim0.5$. The width of the gap and the
tightness of the main sequence indicate low differential extinction in the
field. In Field~2, the stars with $0.5<(U-V)<2.8$ and 
brighter than $V\la22.5$ have no corresponding main sequence stars 
and thus cannot be young stars
migrating from blue to red during their core-He burning phase, 
as is the case for most of the objects in Field~1 with colors 
$0.5<(U-V)<2.5$. These stars in Field~2 are most probably the 
remaining foreground contamination, indicating that the foreground
contamination may affect the numbers of blue and red HeB stars in Field~1
(see next paragraph). Only a few stars are lying along the isochrone of
log(age$)=7.5$ (Fig.~\ref{UVcmd_clean}), while
most of them are consistent with much older ages. 
We conclude that there are no stars younger 
than $\sim40$ Myr in Field~2.

Note that isochrones for metallicities higher than $Z=0.004$ extend 
on the red supergiant edge to
redder values of $U-V$ than the reddest stars in
Field~1 and thus do not fit well our
observations. The ratio of blue to red supergiants strongly depends on
metallicity (Langer \& Maeder~\cite{lm95}, Maeder \& Meynet~\cite{mm01}) and
in principle could be used to constrain the metallicity of the youngest
population in NGC~5128. Counting the number of blue and red supergiants for
stars more massive than $\sim12$ M$_\odot$, we find their ratio to be 
$B/R<0.7-0.8$ (although with high uncertainty due to the possible 
foreground contamination),
in good agreement with the observed $B/R$ value in SMC cluster NGC~300 (see
discussion by Langer \& Maeder~\cite{lm95}). 
The metallicity of $Z=0.004$ (corresponding to [Fe/H]$=-0.7$ dex) is
appropriate for the SMC. However, since the $B/R$ value depends also on other
parameters such as stellar mass, rotation and degree of overshooting
(Maeder \& Meynet~\cite{mm01}), a more detailed comparison with models and
the determination of the metallicity of this youngest stellar population in
NGC~5128 is warranted (Rejkuba et al., in preparation).

The low metallicity implied by the fit of the isochrones on $UV$ CMDs probably
reflects the metallicity of the gas left in the halo of NGC~5128 by the
accreted satellite. Atomic H{\sc i} (Schiminovich et al.~\cite{schiminovich}) 
and molecular CO gas (Charmandaris et al.~\cite{charmandaris}) present
in Field~1 are slightly offset from the position of the diffuse stellar
shell, the obvious remnant from the accreted galaxy. 

\subsection{Optical--Near IR CMDs}
\label{CMD_VK}

\begin{figure*}
\centering
\includegraphics[width=12.8cm,angle=0]{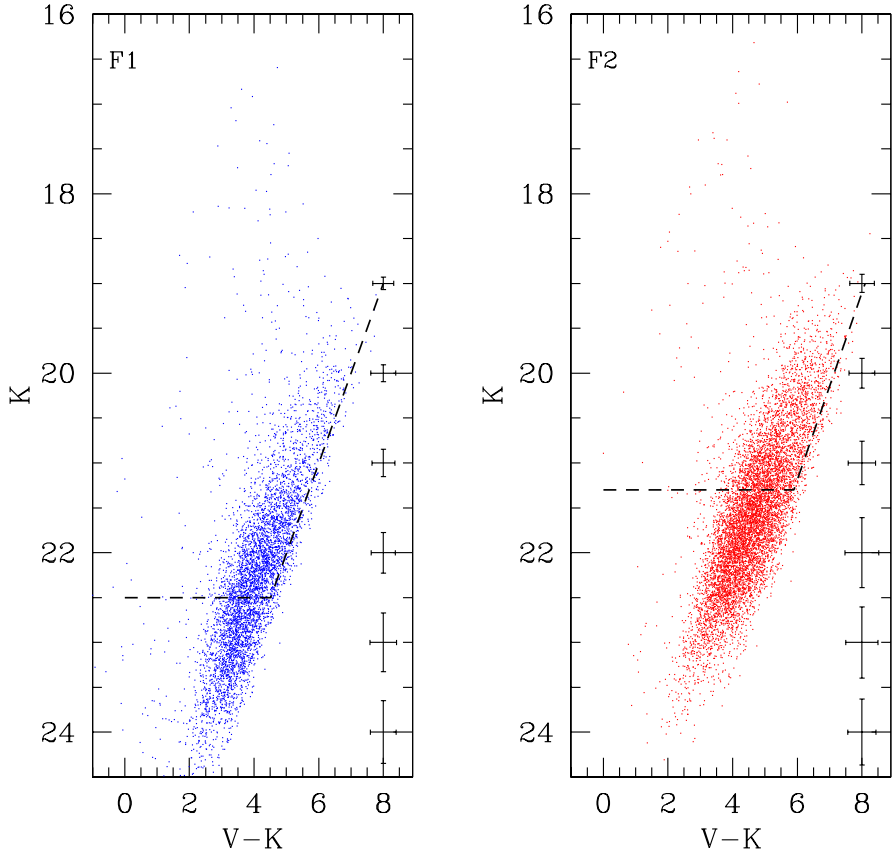}   
  \caption[]{$VK$ color-magnitude diagram for Field~1 (left) and Field~2
(right). The dashed line identifies the 50\% completeness level. }
  \label{VKcmd}
\end{figure*}

\begin{figure*}
\centering
\includegraphics[width=12.8cm,angle=0]{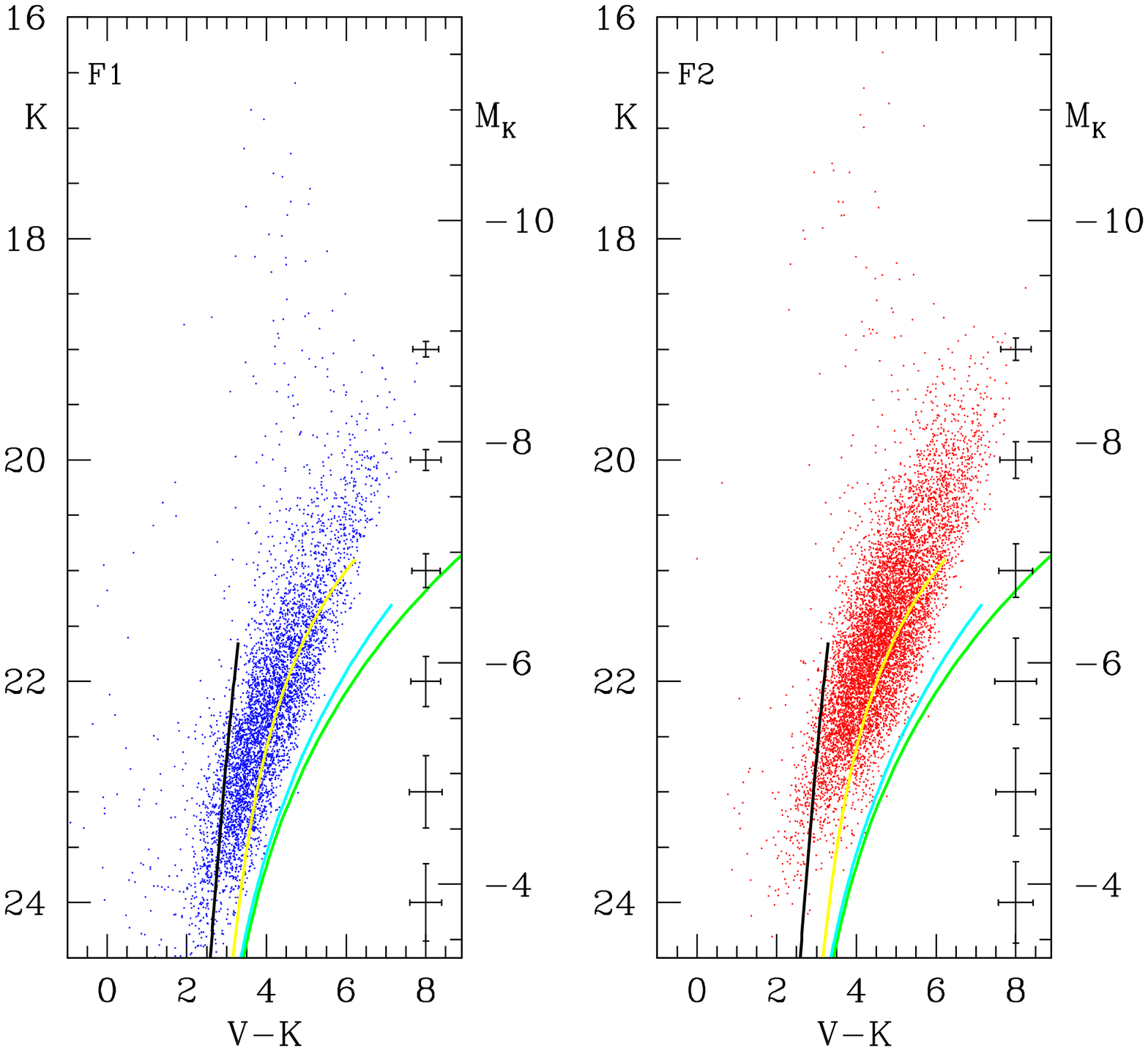}
  \caption[]{$VK$ CMDs with foreground stars statistically subtracted.
Overplotted are fiducial RGB sequences of Galactic globular clusters
(from left to right): M15, 47 Tuc, NGC 6553 and NGC 6528,
from Ferraro et al.~(\cite{ferraro}).}
  \label{VKcmd_clean}
\end{figure*}

Optical-near IR CMDs probe old and intermediate-age
stellar populations. Theoretically, 
more than two thirds of the light in $K$-band is dominated by cool stars on
the red giant branch (RGB) and asymptotic giant branch 
(AGB), and by red dwarfs (Buzzoni~\cite{buzzoni}). The red dwarfs are
too faint to be detected at the distance of NGC~5128, and thus our $VK$ CMDs
are entirely dominated by RGB and AGB stars (Fig.~\ref{VKcmd}).

In Fig.~\ref{VKcmd_clean} we
show $VK$ CMDs of Fields~1 and 2 after the
statistical subtraction of foreground stars. Overlaid are fiducial RGB
sequences of Galactic globular clusters (from left to right: M15, 47 Tuc,
NGC 6553 and NGC 6528; Ferraro et al.~\cite{ferraro})  
spanning a large range of metallicities
($-2.17\le$[Fe/H]$\le-0.23$ dex). As before, we used
a distance modulus of 27.8 and reddening corresponding to $E(B-V)=0.1$
($E(V-K)=0.274$ and $A_K=0.0347$; Rieke \& Lebofsky~\cite{rl85}) 
to adjust the magnitudes and colors of RGB fiducials to those of NGC~5128.
Obviously, most of the stars in Fig.~\ref{VKcmd_clean} belong to the RGB\@. The
right edge of the RGB is quite sharp, with most of the stars being more 
metal-poor
than 47 Tuc ([Fe/H]$=-0.71$ dex) and none appearing to be as metal-rich
as NGC 6553 ([Fe/H]$=-0.29$ dex). However, the latter is due to
incompleteness in $V$-band photometry. 

\begin{figure}
\centering
\includegraphics[width=8.8cm,angle=0]{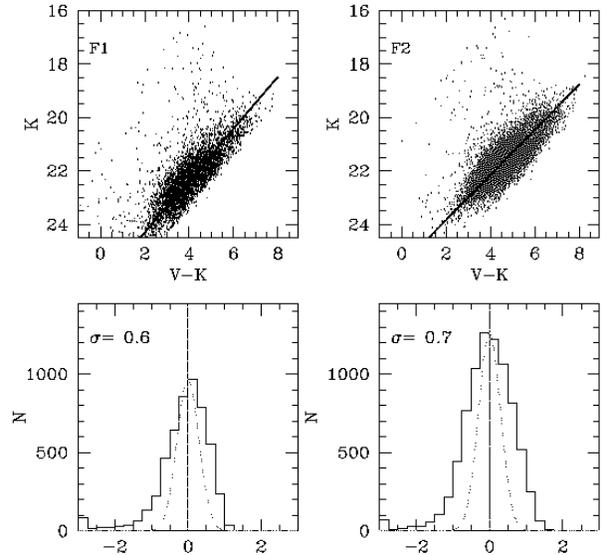}
  \caption[]{The histogram of the observed color distribution along 
the RGB (lower panels) is plotted accounting for the actual slope of the RGB
(thick line in upper panels) in Field~1 (left) and 2 (right). 
The 1$\sigma$ spread in color of the RGB is indicated in the upper left
corner. Overplotted with the dashed line is a Gaussian
with the $\sigma=0.3$, typical of the distribution from the 
photometry errors.}
  \label{colordistrib}
\end{figure}

The spread in color of the 
RGB is larger than the photometric uncertainties (Fig.~\ref{colordistrib}),
indicating the presence of spread in metallicity and/or age. 
The most metal-poor stars have metallicities of $-2$ dex if their ages
correspond to those of Galactic globular clusters. The population we probe
with $VK$ photometry is more metal-poor than $-0.7$ dex. Our $V$-band
images are not deep enough to detect more metal rich giants. 
Walsh et al.~(\cite{walsh}) measured the mean oxygen abundance of five
planetary nebulae in NGC~5128 to be [O/H]$=-0.5 \pm 0.3$ dex, consistent with
the presence of the large population of stars with metallicities below
solar, as we observe in the $VK$ CMDs.

There are 1830 stars in Field~1 and 1197 in Field~2 which have good
$K$-band photometry ($\sigma_K<0.5$,  $-0.7<sharp<0.7$ and $\chi<2.0$) above
the respective $K$-band completeness limits, but which
have not been detected in optical bands. They are uniformly distributed over
the whole ISAAC images.
The large number of very red stars with good photometry in $K_s$ and no 
counterpart in $V$-band suggests that stars more metal-rich than 
[Fe/H]$=-0.7$ dex are present, 
as expected for a luminous giant elliptical galaxy. 
This is in good agreement with the results of 
Harris et al.~(\cite{harris99}) and Harris \& Harris~(\cite{harris00}).

%
%

\section{The Color-Color Diagram}
\label{CCD}

\begin{figure}
\centering
\includegraphics[width=8.8cm,angle=0]{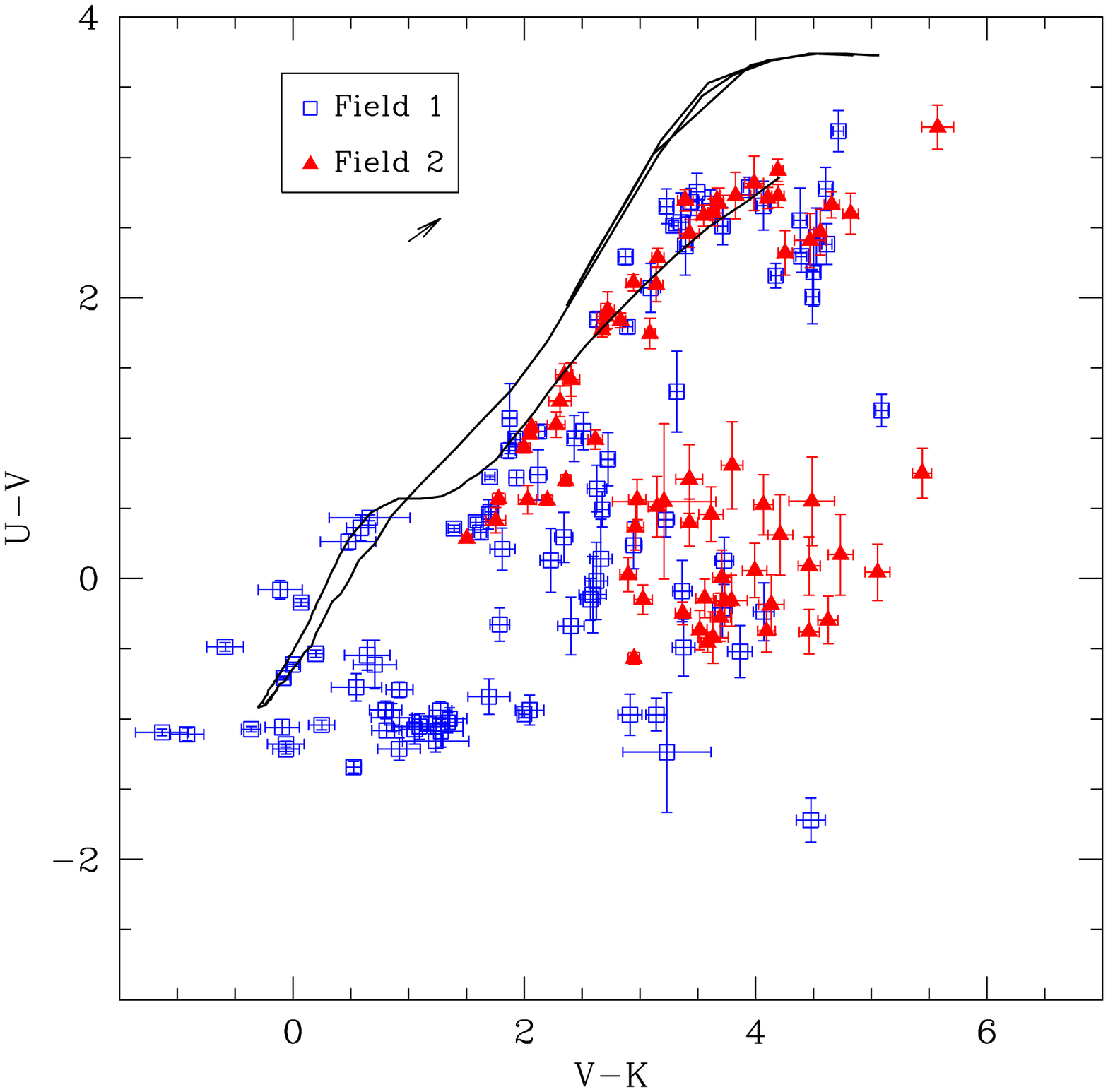}   
  \caption[]{Color-color diagram for the objects matched in $U$, $V$ and
$K_s$ frames that had ALLFRAME photometry uncertainties $<0.5$ mag and 
magnitudes brighter than the 50\% completeness limits in all bands.
Field~1 objects are plotted with open squares and Field~2 with
filled triangles. Overplotted is the Padova isochrone 
(Bertelli et al.~\cite{bertelli94}) for the solar metallicity and log(age)=7.0 yr 
that indicates the locus of the foreground stellar sequence.
Reddening vector has dimensions corresponding to $E(B-V)=0.1$.}
  \label{ccdUVK}
\end{figure}

Fig.~\ref{ccdUVK} shows the color-color diagram for the objects detected
in $U$-, $V$- and $K$-band in Field~1 (open squares) and Field~2 
(filled triangles) that
had ALLFRAME photometry uncertainties $<0.5$ mag and magnitudes brighter
than the 50\%
completeness limits in all bands. The error-bars plotted are the ones given
by ALLFRAME\@. 
A total of 99 objects in Field~1 and 90 in Field~2
satisfy these criteria. Most of them are located along the stellar sequence
that crosses the  color-color diagram in diagonal. Matching the UV and VK
catalogues after the statistical subtraction of the foreground
contamination, the stellar sequence disappears and only 
54 and 44 objects with good photometry in all
three filters are left in Field~1 and 2, respectively.

We use the color-color diagram presented in Fig.~\ref{ccdUVK}
to measure the reddening and the contamination by background galaxies.

\subsection{Reddening}

The objects crossing the color-color diagram in diagonal from top right to
bottom left belong to the Milky Way.  The quality of the fit of the stellar 
sequence with isochrones depends on the age of the isochrone with respect to stars. 
Because the foreground Milky Way stars span
a range of ages, isochrones of different ages fit different
parts of the stellar sequence. As an example, in order to guide the eye we 
overplotted the stellar loci of solar metallicity stars and log(age)=7.0 yr 
(Bertelli et al.~\cite{bertelli94}).
The tight stellar sequence in the $U-V$ vs.\ $V-K$ color-color diagram shows
that the foreground reddening in the two fields is the same. 
Its magnitude was measured by fitting the
theoretical isochrones and it amounts to 
$E(B-V)=0.15\pm0.05$ mag. This is in
excellent agreement with the Schlegel et al. (\cite{schlegel}) extinction
maps and the reddening determined by Fasset \&
Graham~(\cite{fg00}), who used
$UBV$ color-color diagrams to measure a value of
$E(B-V)=0.14\pm 0.02$ mag. 

\subsection{Star-galaxy separation}

Using the color-color plots it is possible to distinguish the foreground
dwarf stars from the most compact unresolved galaxies. 
Galaxies are located in the lower right part of
the diagram in Fig.~\ref{ccdUVK}. The resolved galaxies are not plotted here
since they got rejected by our selection of shape 
parameters in FIND and ALLFRAME. 
Note that the group of objects with $-0.5<(V-K)<2$  are the young stars 
detected in Field~1.
%
%
\section{Jet aligned Star Formation in the Halo}
\label{SF}

\begin{figure}
\centering
\includegraphics[width=8.8cm]{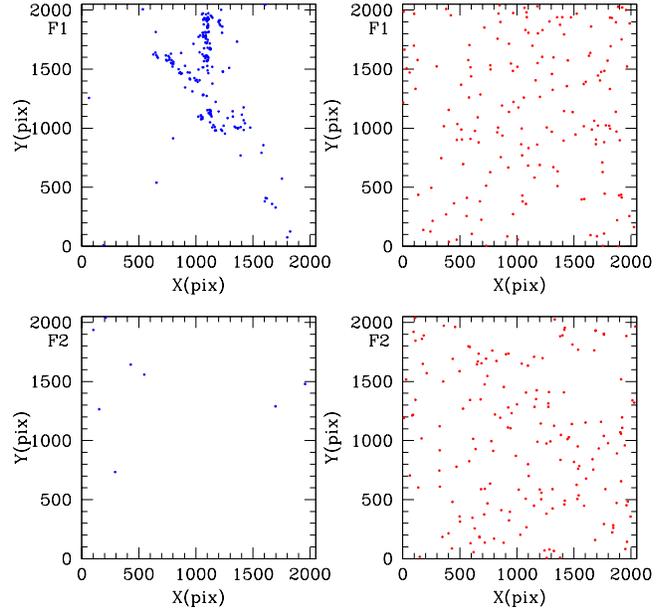}   
  \caption[]{The spatial distribution of the bluest $(U-V)<-0.75$ 
(left panels) and the reddest $(U-V)>2$ (right panels) stars in Field~1 
(top panels)  and 2 (bottom panels) for stars with magnitudes above the 50\%
completeness limit ($V<25$ and $U<25$). The north is 
up and east is to the left. }     
  \label{blue_red}
\end{figure}

The chains of bright blue compact objects and optical filaments, 
first noticed by Blanco et al.~(\cite{blanco}), were recognized 
as young blue supergiants and OB associations 
by Graham \& Price~(\cite{gp81}) and Graham~(\cite{graham98}). 
Our Field~1, situated $\sim14$ kpc away from the center
of the galaxy and coinciding with the NE radio lobe, 
contains part of this ``string of blue knots and filaments.''

In Fig.~\ref{blue_red} we show the spatial distribution of the bluest 
(left) and the reddest (right) stars in both our 
fields. The red stars, belonging to the foreground Galactic 
population, are uniformly distributed in both fields. Field~2
has almost no very young blue stars, while the majority of the young 
population in Field~1 is aligned with the jet direction coming from the 
nucleus of the galaxy. The vertical structure to the right of the jet is 
located at the leftmost edge of a large H{\sc i} cloud found by Schiminovich 
et al.~(\cite{schiminovich}). In the northern part of this field, 
a CO molecular cloud was  recently discovered by Charmandaris 
et al.~(\cite{charmandaris}).

Mould et al.~(\cite{mould00}), Graham~(\cite{graham98}) and Fasset
\& Graham~(\cite{fg00}) report on the presence of the 
collimated star formation in the halo of NGC~5128. 
They favor the scenario in which a past interaction between the radio jet
(Morganti et al.~\cite{morganti99}) and the H{\sc i} cloud 
complex (Schiminovich
et al.~\cite{schiminovich}) is responsible for the star formation. 

Finding active star formation far out in the
halo of an elliptical galaxy is unusual. Understanding the triggering
mechanism and the origin of the star formation, as well as the nature of the
underlying ionized gas (Graham~\cite{graham98}, Morganti et 
al.~\cite{morganti91,morganti92}),
is important for explaining the filamentary
emission and jet-aligned structures observed in other galaxies 
(e.g., in NGC~1275: Lazareff et al.~\cite{lazareff}, 
Sabra et al.~\cite{sabra};
in M87: Gavazzi et al.~\cite{gavazzi}; and in other more distant 
radio sources: Rees~\cite{rees}, Best et al.~\cite{bestetal}, De Breuck et
al.~\cite{debreuck}).

The alignment of the luminous blue stars with the radio axis of 
NGC~5128 (Fig.~\ref{blue_red}) may shed some light on the debated issue of
the observed alignment of the radio and optical structures in high-$z$
radio galaxies (Rees~\cite{rees}, Best et al.~\cite{bestetal}). One of the
proposed alignment models is based on the assumption that the action of the
jets tunneling through a galactic medium affects the gas via shock heating,
thereby inducing vigorous star formation (Begelman \&
Cioffi~\cite{begelman}, Rees~\cite{rees}, Daly~\cite{daly}). In particular,
as Rees~(\cite{rees}) discussed in detail, the stars form from the
cool-phase gas (with $T\leq10^4$ K) in clouds massive or dense enough to be
Jeans-unstable.

This mechanism may work well for massive high-$z$ radio galaxies, in which
most of the baryonic mass would still be gaseous. In the present day giant
ellipticals, on the other hand, little gas is usually left to support
strong star formation.  However, probably due to previous accretion of a
gas rich galaxy, NGC~5128 has substantial cool, dense clouds of molecular
CO and atomic H{\sc i} gas in its halo (Schiminovich et
al.~\cite{schiminovich}, Charmandaris et al.~\cite{charmandaris}). We note
that these clouds could form the clumpy ISM/IGM necessary for the `bursting
bubble' model proposed by Morganti et al.~(\cite{morganti99}).

We have shown clear evidence for blue stars as young as $10\times10^6$ yr
(Fig.~\ref{UVcmd_clean}).  We also see underlying emission, particularly in
our $U$-band images, which probably comes from the [OII]~3727 emission
line.  High-resolution spectra with good signal-to-noise are needed to
decide whether this emission is due to gas being photoionized by the newly
formed stars or being photoionized by particles coming directly from the
nucleus.

Other low-redshift galaxies also show filamentary ionized gas
structures in their halos. In the case of M87 (Gavazzi et 
al.~\cite{gavazzi}), the filament of ionized gas in the NE 
part of the halo coincides
with the Eastern radio-lobe, in much the same way as in
the much closer NGC~5128. It would be interesting to search for 
young stars associated with that filament too.

%
%
\section{Intermediate Age Population}
\label{AGB}

\begin{figure*}
\centering
\includegraphics[width=12.8cm,angle=0]{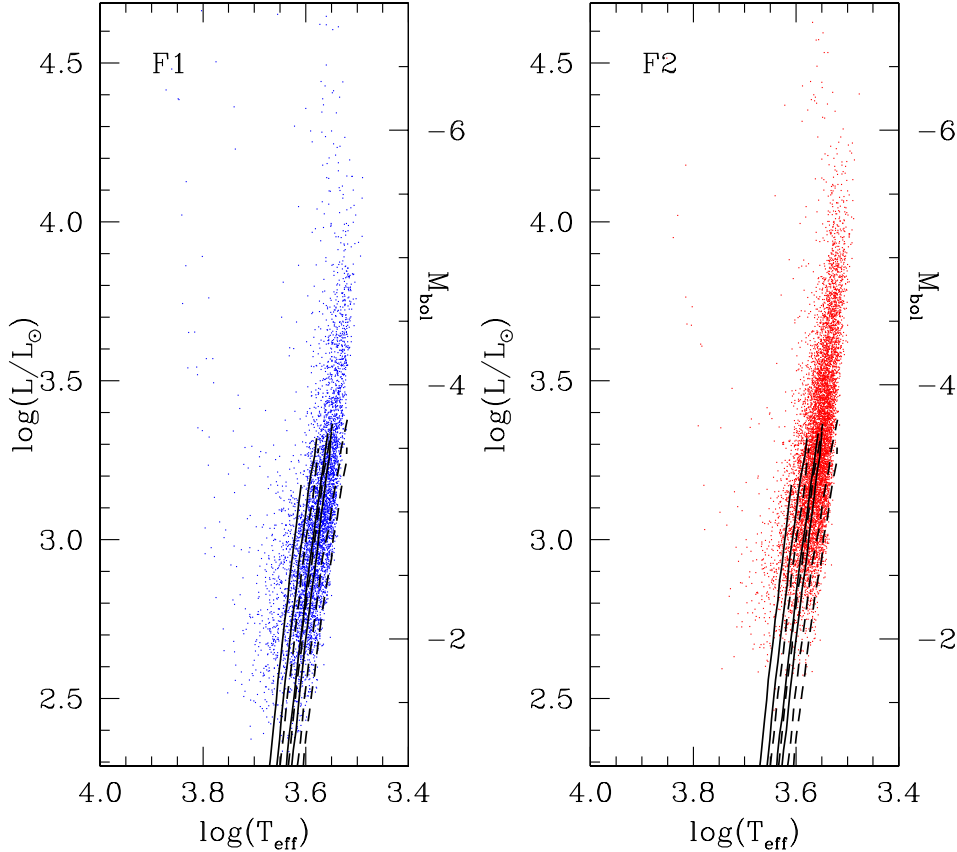}   
  \caption[]{The HR diagram for old and intermediate stellar populations in
Field~1 and 2. Overplotted are evolutionary tracks from Padova (Girardi et
al.~\cite{g00}) for stars with masses M=0.6, 0.8, 1.2 and 1.6 M$_\odot$ and
$Z=0.004$ (full line) and $Z=0.008$ (dashed line).
}     
  \label{HRD}
\end{figure*}

In contrast with the conclusions of HST studies 
(Harris et al.~\cite{harris99}, Harris \& Harris~\cite{harris00}) of stellar
populations in NGC~5128, we detect not only old population II stars, but also
a significant number of stars with magnitudes brighter than the tip of
RGB ($M_K^{\rm Tip} = -0.64(\pm0.12){\rm [M/H]} - 6.93(\pm0.14)$; Ferraro et
al.~\cite{ferraro}). 

We took bolometric corrections from Bessell \& Wood
(\cite{bw84}) and the empirical fit of ($V-K$) vs.\ $T_{\rm eff}$ from 
Bessell, Castelli \& Plez (\cite{bcp})
to transform our $K$--($V-K$) CMDs to the theoretical plane
(Fig.~\ref{HRD}). Overplotted on the H-R diagrams are 
the Padova tracks  from Girardi et al.~(\cite{g00}) 
for the first-ascent giant branch stars
with masses M=0.6, 0.8, 1.2 and 1.6 M$_\odot$  and 
metallicities $Z=0.004$ (full lines) and $Z=0.008$ (dashed lines). The sharp
cut-off on the right side of the H-R diagrams is due to incompleteness 
in $V$-band.
 
In discussing the H-R diagram, we should consider blending.
According to theoretical predictions (Renzini~\cite{renzini}), 
at the fiducial galactocentric distances of 9, 12.8 and 17.9 kpc, 
the number of
blends consisting of 2 stars belonging to the tip of RGB is $\sim
1300$, 300 and  55 stars 
(per $2.2\arcmin \times 2.2\arcmin$; i.e. field of view of ISAAC),
respectively. 
In this calculation, the $B$-band
surface brightness measurements from Mathieu, Dejonghe \& Hui (\cite{MDH})
were used. Obviously, the inner most regions of the galaxy, at the distance
$<10$ kpc from the nucleus are too crowded to give accurate
photometry for RGB stars and most of the stars above the tip of RGB. 

We attempted to use the NIC3 HST images overlapping with our Field~2
(galactocentric distance $R_{gc}\sim9$ kpc) data to
improve the resolution. However, we found out that 
many faint
stars visible on ISAAC images are completely within the noise of NIC3
data. The resolution of our ISAAC data in the best seeing is as good as 
that of
NICMOS images, with the advantage of having much better S/N and a more 
stable PSF\@.
The confirmation of the number of AGB stars in Field~2 can be
obtained only through the analysis of the LPVs (Rejkuba et al.
in preparation). 

\begin{figure}
\centering
\includegraphics[width=8.8cm]{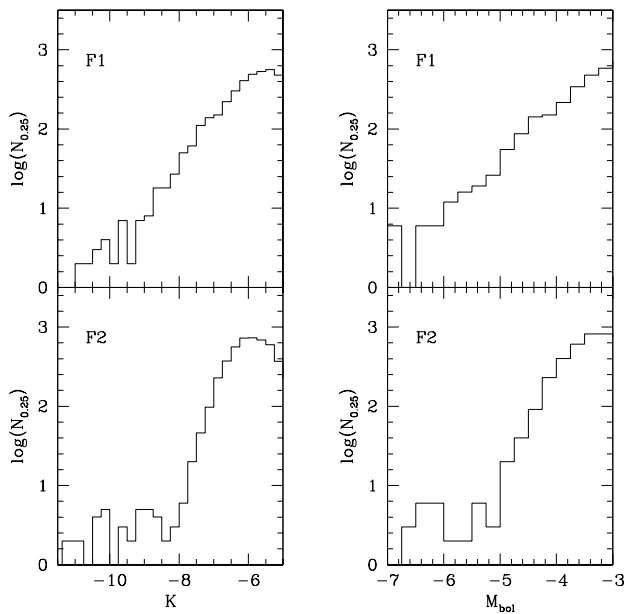}   
  \caption[]{$K$-band (left) and bolometric magnitude luminosity functions
(right) for stars
redder than $(V-K_s)>2$ in Fields~1 and~2.
}     
  \label{LF}
\end{figure}

In Field~1 ($R_{gc}\sim14$ kpc), our outer shell field, 
crowding is not that severe. The expected number of two-star blends
at the RGB tip in
this field ranges from 300--50, as the surface brightness drops across the
field.  The total number of such blends is therefore less 
than $\sim 200$ in the most pessimistic calculation. 
On the other hand, the number of stars above the tip of RGB
($M_{\rm bol}\sim-3.7$; Girardi et al.~\cite{g00}) is
768. Subtracting the number of possible blends and allowing for a few of the
brightest and bluest stars ($\sim 30$ stars with $M_K<9$) 
to be the remaining foreground contamination,
there are still more than 500 stars whose position in the H-R diagram 
and CMD is
consistent with an intermediate-age AGB population. 

In the inner halo field, 
Field~2, the number of stars
above the tip of RGB is 2844. After subtracting the number of expected
blends ($\la 1500$), the number of AGB stars is 2.5 times larger than in
Field~1. This confirms the presence of gradients in the intermediate-age
population within the halo of NGC~5128, as suggested by Marleau et
al.~(\cite{marleau00}).

The intermediate-age AGB 
 population could have easily been missed in $V$ and $I$-band HST 
studies due to small field of view and small ($<0.5$ mag) 
optical magnitude difference between the tip of
the RGB and the tip of AGB. The AGB stars are up to $\sim 2$ magnitudes 
brighter in $M_K$ and $M_{\rm bol}$ than RGB stars and thus are easily 
detectable in
near-IR. Thanks to this, Marleau et al.~(\cite{marleau00}) could detect some
AGB stars in the much smaller NICMOS field. 

The brightest stars in M32, in the Galactic bulge and the bulge of M31 have
similar brightnesses, reaching $M_{\rm bol} = -5.5$
(Freedman~\cite{freedman92}, Elston \& Silva~\cite{es92},
Frogel \& Whitford~\cite{fw87}, Rich \& Mould~\cite{rm91}). Due to this
similarity,  
Davidge \& van den Bergh~(\cite{dvdb01}) suggested that the tip of AGB could
be used as standard candle for determination of distances.
In Fig.~\ref{LF} we present the $K_s$-band and
bolometric magnitude luminosity functions. The tip of the AGB is observed at
bolometric magnitude of $-5$ in both fields (in spite of the crowding in
Field~2), consistent with the adopted distance modulus for NGC~5128. 
Most of the stars brighter than this magnitude (as well as stars
brighter than $M_K = -9$; see left panels in Fig.~\ref{LF}) probably
belong to our Galaxy.

%
%
\section{The Shell}
\label{shell}

\begin{figure*}
\centering
\includegraphics[width=12.8cm]{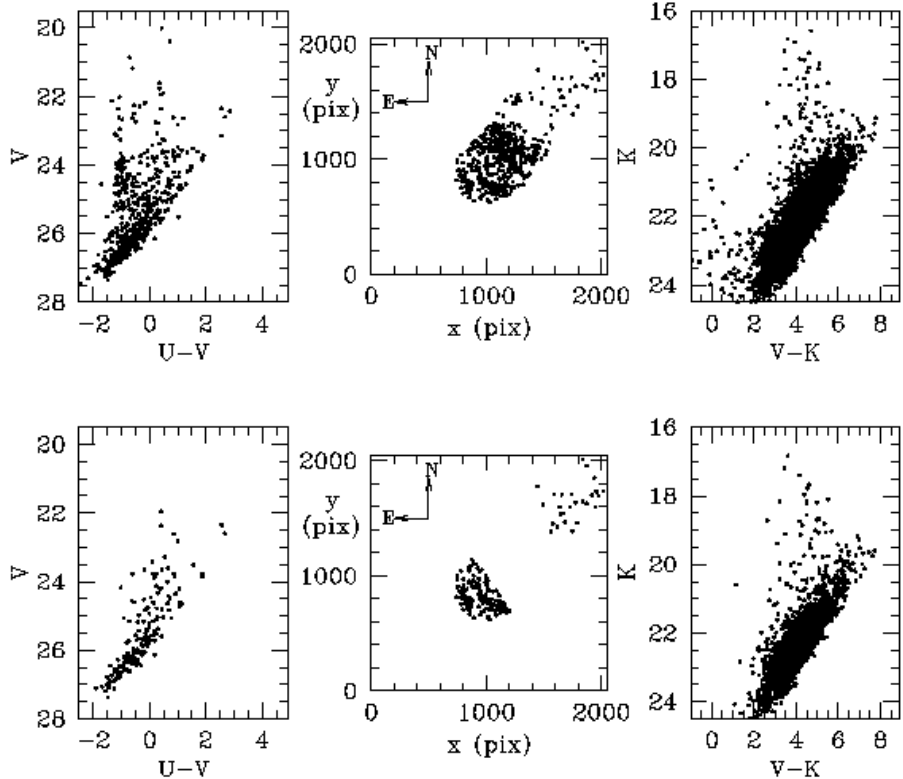}   
  \caption[]{The color-magnitude diagrams for the stars located at the
position of the diffuse NE shell. }
  \label{shellfig}
\end{figure*}

In our deep $V$-band image of Field~1, the prominent NE shell is easily
recognizable as the region with higher surface brightness, and also with higher
density of stellar objects.
All the stars that are spatially coincident with the shell 
on the $V$-band image are shown in
Fig.~\ref{shellfig}. In the left two panels we plot the 
$UV$ CMD of all stars coincident with the position of the diffuse shell
(middle panel), while
the corresponding $VK$ CMD is on the right. 
The much smaller
field of view of ISAAC is centered on the shell so that the shell stars
cover most of the image. Note that projected on the shell is the stream of
blue, young objects that are aligned with the jet direction. Excluding the
stars that are coincident with the star forming area (lower panels), 
the stellar
population belonging to the shell is very similar to the one observed in
the halo of NGC~5128, i.e., our Field~2 population (Fig.~\ref{UVcmd}, 
\ref{VKcmd}).
As in the rest of the galaxy there is a significant spread in metallicity.

The presence of stellar shells or ripples around early-type galaxies is
usually explained as evidence of the accretion and later merging of a
smaller companion galaxy. The simulations of the stellar component in the
merging galaxies have shown that the shells are created as a result of
phase-wrapping or of spatial-wrapping of the tidal debris
(Quinn~\cite{quinn}; Hernquist \& Quinn~\cite{hernquist&quinn89}).
The companion galaxies may contain 
different stellar populations with respect to the giant elliptical 
(e.g. most dwarf galaxies have much lower metallicity than
giants Mateo~\cite{mateo}).

Here for the first time we obtained the
magnitudes and colors of stars belonging to a shell. They are
surprisingly similar to the rest of the halo stars in NGC~5128. However,
due to incompleteness of our $V$-band photometry,
we cannot probe the most metal rich part of the halo of NGC~5128. 
Very high resolution and deeper observations of the shell stars are 
possible with the Advanced Camera for Surveys at Hubble Space Telescope.
In order to be able to better separate the stars that genuinely 
belong to shells from the ``normal'' halo stars in NGC~5128 and 
confirm the present result a more quantitative comparison between the 
stellar populations in our two fields is necessary 
(Rejkuba et al., in preparation). 

%
%

\section{Summary}
\label{summary}

Using the VLT with FORS1 and ISAAC, 
we have resolved stars in the halo and in the diffuse north-eastern shell of
the closest giant elliptical galaxy, NGC~5128. With the $U$, $V$ and $K_s$-band
photometry we probe different stellar populations. The $U$ filter, in
particular, is sensitive to the youngest stars. It has revealed the striking
difference in the star formation history between the 
the inner halo field (Field~2) and outer halo field (Field~1), the latter
coinciding with the diffuse shell that is presumably the signature of a
recent merger.

In Field~1, stars as young as 10 Myr are present,
while there are no stars younger than at least 40 Myr in Field~2. 
Thanks to the high quality of our $UV$ photometry, we also detect the gap
between the main sequence and the blue core-helium burning supergiants. The
foreground and background 
contamination has been corrected using the Besan\c{c}on Galaxy
model and our own color-color diagram. 

The recent star formation in Field~1 is approximately aligned with the
direction of the jet coming from the center of AGN. However, there is
also a chain of young blue stars offset from the jet direction that is
aligned with the edge of the molecular clouds. The presence of the 
atomic and molecular gas
(Schiminovich et al.~\cite{schiminovich}, Charmandaris et
al.~\cite{charmandaris}) and their association 
with the diffuse stellar shell is
compatible with the dynamical scenario of phase-wrapping following the
merger of a smaller gas-rich galaxy with NGC~5128 (Quinn~\cite{quinn}). 
In this scenario, the	
interaction of the material coming from the AGN in the center of the galaxy
with the material left from the merger stimulated the star
formation in the halo,  $\sim15$ kpc away from
the galactic centre. The metallicity of the newly formed stars gives an upper
limit on the abundance of the gas of the accreted satellite with value of
$Z=0.004$, a value typical of the SMC\@. 

While the $UV$ CMDs probed the youngest stellar populations, the combination
of optical and near-IR filters is most sensitive to old and intermediate-age
populations. In our $VK$ CMDs we observe a very wide and extended giant
branch. The width indicates the presence of Population~II stars with
metallicities ranging from $-2$ to $-0.7$ dex. Unfortunately, our $V$-band
images are not deep enough to detect more metal rich stars, but the large
number of stars detected on $K$-band images without a counterpart in $V$-band
suggests an even redder and more metal-rich RGB population, in agreement with
the previous HST studies (Soria et al.~\cite{soria}, 
Harris et al.~\cite{harris99}, Harris \& Harris~\cite{harris00}). 

In contrast with the previous two studies, we find a very extended 
giant branch up to $M_{\rm bol}=-5$, which 
reveals the presence of the intermediate-age AGB population. Our extensive
crowding experiments and theoretical predictions of the amount of blending
(Renzini~\cite{renzini}) demonstrate that blending due to crowding 
is not significant enough to mimic this population in the outer
field (Field~1). In the inner field, the extent and the properties of the AGB
population will be assessed through the study of the long period
variables that have been detected in both fields (Rejkuba et al. in
preparation).

\begin{acknowledgements}

We thank Manuela Zoccali for help with DAOPHOT and ALLFRAME and Joel Vernet
for help with IDL. We thank Francesco Ferraro who 
kindly provided the tables of VK
globular cluster fiducial RGB sequences in electronic form. Felipe
Barrientos kindly provided the U$-$V vs. V$-$K colors for galaxies of
different morphological types and redshifts.
MR acknowledges ESO studentship programme.  TRB is grateful to the
Australian Research Council and P. Universidad Cat\'olica for financial
support and to the DITAC International Science \& Technology Program. This
work was supported by NASA Grant GO-07874.01-96A and by the Chilean Fondecyt
No.\ 01990440 and 799884. Finally, we are grateful to the referee Mario
Mateo for his helpful and thorough report.
\end{acknowledgements}

\end{document}